\documentclass{aa}
\usepackage{graphics}

\begin{document}

   \thesaurus{01     
              (02.02.1;  
               13.07.3;  
               13.07.1;  
               13.07.2)}  
   \title{On Evolution of the Pair-Electromagnetic Pulse of a Charge Black Hole}


   \author{Remo Ruffini
          \inst{1}
          \and
          Jay D. Salmonson\inst{2}
	    \and
	    James R. Wilson\inst{2}
	    \and          
	    She-Sheng Xue\inst{1}
          }

   \institute{I.C.R.A.-International Center for Relativistic Astrophysics
and
Physics Department, University of Rome ``La Sapienza", I-00185 Rome,
Italy\\
	\and Lawrence Livemore National Laboratory, University of
     California, Livermore, California, U.S.A.
             }


   \maketitle

   \begin{abstract}

Using hydrodynamic computer codes, we study the possible patterns of
relativistic expansion of an enormous pair-electromagnetic-pulse
(P.E.M. pulse); a hot, high density plasma composed of photons,
electron-positron pairs and baryons deposited near a charged black
hole (EMBH). On the bases of baryon-loading and energy conservation, we
study the bulk Lorentz factor of expansion of the P.E.M. pulse by
both numerical and analytical methods.

\keywords{black holes -- gamma ray bursts
               
               }
   \end{abstract}

%

In the paper by Preparata, Ruffini \& Xue (1998),
the ``dyadosphere" is defined as the 
region outside the horizon of a
EMBH where the electric field exceeds the critical value for $e^+ e^-$
pair production. In Reissner-Nordstrom EMBHs, the horizon 
radius is expressed as
\begin{equation}
r_{+}={GM\over c^2}\left[1+\sqrt{1-{Q^2\over GM^2}}\right].
\label{r+}
\end{equation}
The outer limit of the dyadosphere is
defined as the radius $r_{\rm ds}$ at which the electric field of the
EMBH equals this critical field
\begin{equation}
r_{\rm ds}= \sqrt{{\hbar e Q \over m^2 c^3}}.
\label{rc}
\end{equation} 
The total energy of pairs, converted from the static electric energy,
deposited within a dyadosphere is then
\begin{equation}
E^{\rm tot}_{e^+e^-}={1\over2}{Q^2\over r_+}(1-{r_+\over r_{\rm ds}})(1-
\left({r_+\over r_{\rm ds}}\right)^2) ~.
\label{tee}
\end{equation}
In Wilson (1975, 1977) a black hole charge of the order $10 \%$ was
formed. Thus, we henceforth assume a black hole charge $Q = 0.1
Q_{max},Q_{max}=\sqrt{G}M$ for our detailed numerical calculations.  
The range of energy is of interest as a possible gamma-ray burst source.

In order to model the radially resolved evolution of the energy
deposited within the $e^+e^-$-pair and photon plasma fluid created
in the dyadosphere of EMBH, we need to discuss the
relativistic hydrodynamic equations describing such evolution.

The metric for a Reissner-Nordstrom black hole is
\begin{equation}
d^2s=-g_{tt}(r)d^2t+g_{rr}(r)d^2r+r^2d^2\theta +r^2\sin^2\theta
d^2\phi ~,
\label{sw}
\end{equation}
where $g_{tt}(r)=-g^{-1}_{rr}(r)= - \left[1-{2GM\over c^2r}+{Q^2G\over
c^4r^2}\right]$.

We assume the plasma fluid of $e^+e^-$-pairs, photons and baryons to be a simple 
perfect fluid in the curved spacetime (Equation
\ref{sw}). The stress-energy tensor describing such an fluid is given
by (\cite{mtw})
\begin{equation}
T^{\mu\nu}=pg^{\mu\nu}+(p+\rho)U^\mu U^\nu
\label{tensor}
\end{equation}
where $\rho$ and $p$ are respectively the total proper energy density
and pressure in the comoving frame. The $U^\mu$ is the four-velocity
of the plasma fluid. The baryon-number and energy-momentum conservation 
laws are 
\begin{eqnarray}
(n_B U^\mu)_{;\mu}&=&(n_BU^t)_{,t}+{1\over r^2}(r^2 n_BU^r)_{,r}= 0~,
\label{contin}\\
(T_\mu^\sigma)_{;\sigma}&=&0,
\label{contine}
\end{eqnarray}
where $n_B$ is the baryon-number density. The radial component of Equation (\ref{contine}) reduces to 
\begin{eqnarray}
&&{\partial p\over\partial r}+{\partial \over\partial t}\left((p+\rho)U^t U_r\right)+{1\over r^2} { \partial
\over \partial r}  \left(r^2(p+\rho)U^r U_r\right)\nonumber\\
&&-{1\over2}(p+\rho)\left[{\partial g_{tt}
 \over\partial r}(U^t)^2+{\partial g_{rr}
 \over\partial r}(U^r)^2\right] =0 ~.
\label{cmom2}
\end{eqnarray}
The component of the energy-momentum conservation equation
(\ref{contine}) along a flow line is
\begin{eqnarray}
&&U_\mu(T^{\mu\nu})_{;\nu}=\nonumber\\
&&(\rho U^t)_{,t}+{1\over r^2}(r^2\rho
U^r)_{,r}+p\left[(U^t)_{,t}+{1\over r^2}(r^2U^r)_{,r}\right]=0 ~.
\label{conse1}
\end{eqnarray}
Equations (\ref{contin}) and (\ref{conse1}) give rise to the relativistic hydrodynamic
equations.

We now turn to the analysis of $e^+e^-$ pairs initially created in
the Dyadosphere. Let $n_{e^\pm}$ be the proper densities of
electrons and positrons($e^\pm$). 
The rate equation for $e^\pm$ is
\begin{equation}
(n_{e^\pm}U^\mu)_{;\mu}=\overline{\sigma v} \left[n_{e^-}(T)n_{e^+}(T) - n_{e^-}n_{e^+}\right] ~,
\label{econtin}
\end{equation}
where $\sigma$ is the mean pair annihilation-creation cross-section,
$v$ is the thermal velocity of $e^\pm$, and $n_{e^\pm}(T)$
are the proper number-densities of $e^\pm$, given by
appropriate Fermi integrals with zero chemical potential. The equilibrium temperature
$T$ is determined by the thermalization processes occuring in the
expanding plasma fluid with a total proper energy-density $\rho$,
governed by the hydrodynamical equations
(\ref{contin},\ref{conse1}). We have
\begin{equation}
\rho = \rho_\gamma + \rho_{e^+}+\rho_{e^-}+\rho^b_{e^-}+\rho_B,
\label{eeq}
\end{equation}
where $\rho_\gamma$ is the photon energy-density and $\rho_{e^\pm}$ is the
proper energy-density of $e^\pm$. In Equation (\ref{eeq}), $\rho^b_{e^-}+\rho_B$ are baryon-matter contributions.  We
can also, analogously, evaluate the total pressure $p$. We define the total proper internal energy density $\epsilon$ and the baryon mass density $\rho_B$ in the comoving frame, and have the equation of state ($\Gamma$ is thermal index)
\begin{equation}
\epsilon \equiv \rho - \rho_B,\hskip0.2cm \rho_B\equiv n_Bmc^2,\hskip0.2cm
\Gamma = 1 + { p\over \epsilon}.
\label{state}
\end{equation} 

The calculation is initiated by depositing the total energy 
(\ref{tee}) between the Reissner-Nordstrom radius $r_+$ and the
dyadosphere radius $r_{ds}$. The calculation is continued as the plasma fluid
expands, cools and the $e^+e^-$ pairs recombine, until it becomes optically
thin:
\begin{equation}
\int_R (n_{\rm pairs}+n^b_e) \sigma_{T} dr = {2 \over 3}\label{ttp}
\end{equation}
where $\sigma_{T}$ is the Thomson cross-section, $n^b_e$ is the number-density of ionized electrons and integration is over the radial size of the expanding plasma fluid in the comoving frame. Here, we only present $n^b_e=0,\rho_B=0$ case. 


   \begin{figure}
   \resizebox{\hsize}{8cm}{\includegraphics{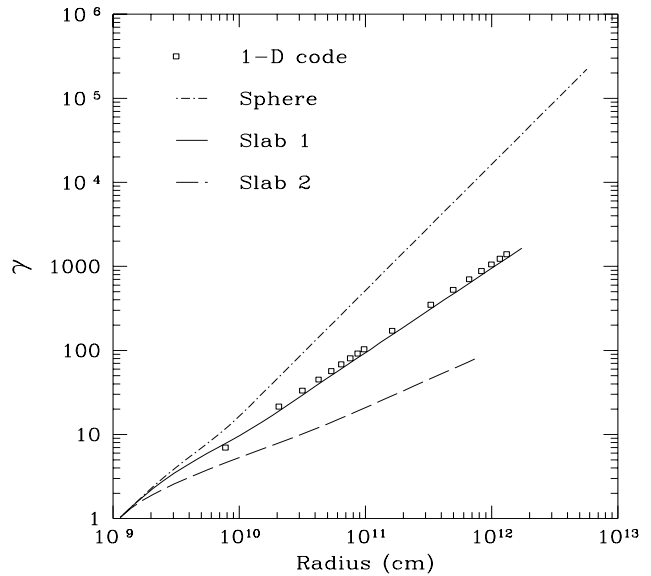}}
      \caption[]{ Lorentz $\gamma$ as a function of radius.
Three models for the expansion pattern of the $e^+e^-$ pair plasma are
compared with the results of the one dimensional hydrodynamic code for
a $1000 M_\odot$ black hole with charge $Q = 0.1 Q_{max}$.  The 1-D
code has an expansion 
pattern that strongly resembles that of a shell
with constant coordinate thickness.
\label{figshells}}
   \end{figure}
%

We use a computer code (Wilson, Salmonson \& Mathews 1997, 1998) to
evolve the spherically symmetric hydrodynamic equations for the
baryons, $e^+e^-$-pairs and photons deposited in the Dyadosphere.  In
addition, we use an analytical model to integrate the spherically
symmetric hydrodynamic equations with the following geometries of
plasma fluid expansion: (i) spherical model: the radial component of
four-velocity $U(r)=U{r\over {\cal R}}$, where $U$ is four-velocity at
the surface (${\cal R}$) of the plasma, (ii) slab 1: $U(r)=U_r={\rm
const.}$, the constant width of expanding slab ${\cal D}= R_\circ$ in
the coordinate frame of the plasma; (iii) slab 2: the constant width
of expanding slab $R_2-R_1=R_\circ$ in the comoving frame of the
plasma. 

We 
compute the relativistic Lorentz factor $\gamma$ of the expanding
$e^+e^-$ pair and photon plasma.
We compare these hydrodynamic calculations with simple models of the
expansion. In Figure (1) we see a comparison of the
Lorentz factor of the expanding fluid as a function of radius for all
of the models.  We can see that the one-dimensional code (only a few 
significant points 
are pesented) matches the
expansion pattern of a shell of constant coordinate thickness (slab 1).

We have shown that a relativistically expanding P.E.M. pulse can
originate from the Dyadosphere of a EMBH. The P.E.M. pulse can produce
gamma-ray bursts having the general characteristics of observed
bursts.  For example, the burst energy for a $1000 M_\odot$ BH is $3
\times 10^{54}$ ergs with a spectral peak at $500$ keV and a pulse
duration of 40 seconds (Ruffini, Salmonson, Wilson \& Xue, 1999).
This oversimplified model is encouraging enough to demand further
study of the Dyadosphere created by EMBHs.

\end{document}